# Does the method of quasi-averages lead to the periodic density in a crystal?

V. A. Golovko

Moscow State University of Mechanical Engineering (MAMI)

Bolshaya Semenovskaya 38, Moscow 107023, Russia

E-mail: fizika.mgvmi@mail.ru

Abstract

Since the Gibbs distribution function always yields a density of particles constant in space, in order to obtain the periodic density characteristic of a crystal it is usual to mention the method of quasi-averages. In the present paper it is shown that the method of quasi-averages does not lead to the periodic density as well as the initial Gibbs distribution. It is also discussed how the crystalline state can be investigated by means of statistical mechanics.

The Gibbs method based on the use of the canonical and other equilibrium ensembles is widely employed in the theory of liquids and gases. However its direct utilization for description of the crystalline state does not yield the desired result. The problem is that the single particle distribution function proportional to the crystal density must be spatially periodic in a crystal. At the same time, if one integrates the Gibbs distribution function over the coordinates of all particles but one, the resulting single particle distribution function will always be constant in space as in the case of the liquids and gases. This can be easily proven.

To this end we write down the canonical Gibbs distribution function

$$F_G = Ce^{-\beta H}, \qquad (1)$$

where $\beta = 1/k_B T$ with the Boltzmann constant $k_B$ and $C$ is a normalization factor. We represent the Hamiltonian $H$ of the system as $H = E_k + E_p$ with

$$E_k = \sum_{i=1}^{N} \frac{mv_i^2}{2}, \qquad E_p = \sum_{i<j}^{N} K(|\mathbf{r}_i - \mathbf{r}_j|). \qquad (2)$$

In this paper it is assumed that $N$ particles of mass $m$ interact via a two-body potential $K(|\mathbf{r}_i - \mathbf{r}_j|)$ and occupy a volume $V$. When integrating the distribution of (1) over the coordinates of all particles but one, e.g. $\mathbf{r}_1$, we make the replacement $\mathbf{r}_i = \mathbf{r}_i' + \mathbf{r}_1$, $i = 2,\ldots,N$. Then the coordinate $\mathbf{r}_1$ disappears from the potential energy $E_p$ and the resulting single particle distribution function will be constant in space at any temperature $T$. It may be remarked that the above replacement shifts the position of the volume of integration $V$ whereas the shift depends on $\mathbf{r}_1$ but the infinite range of integration remains infinite in the thermodynamic limit implied ($V \to \infty$ and $N \to \infty$ with $N/V =$ constant).

To date it is customary to accept that the periodicity of density in a crystal may be obtained with the aid of the procedure called the method of quasi-averages [1–3]. In the case of the crystal the method implies introduction of a periodic external potential field $U_e(\mathbf{r})$ (added to $E_p$) which breaks the symmetry of the system under consideration. Afterwards one should proceed to the thermodynamic limit whereupon one can allow the auxiliary field $U_e$ to vanish. It is believed that the symmetry breaking imposed by the field $U_e$ will be maintained by the system itself. It should be stressed that one has to take the thermodynamic limit first and the limit $U_e \to 0$ after; otherwise no periodic density emerges. There is no strict proof of this procedure and no examples are available as to applying the procedure in actual practice for investigation of crystals. In the present paper we shall not discuss deficiencies and contradictions in the method of quasi-averages (see [4]) but we shall simply demonstrate that the method does not lead to the periodic density.

With regard to (2) we now recast the Gibbs distribution of (1) in the form

$$F_G = C\,\Phi(\mathbf{v})\Psi(\mathbf{r}); \quad \Phi(\mathbf{v}) = e^{-\beta E_k}, \quad \Psi(\mathbf{r}) = e^{-\beta(E_p + U_e)}. \tag{3}$$

Here and henceforth $\mathbf{v}$ denotes the set of velocities $\mathbf{v}_1,\ldots,\mathbf{v}_N$ and $\mathbf{r}$ the set of coordinates $\mathbf{r}_1,\ldots,\mathbf{r}_N$. If we increase $N$, in $\Phi(\mathbf{v})$ there will appear only extra multipliers identical in form according to the appearance of extra summands in the kinetic energy $E_k$ from (2). As to the function $\Psi(\mathbf{r})$, it will change in a more complicated way while solely this function depends on $U_e(\mathbf{r})$. Continuing to increase $N$ we shall proceed to the thermodynamic limit. It is convenient to multiply and divide the right-hand side of (3) by $e^{-\beta E_p}$ conserving the previous designation $E_k + E_p = H$. Upon denoting the function $F_G$ of (3) in the thermodynamic limit as $F$ we obtain [cf. (1)]

$$F = C e^{-\beta H}\,\tilde{\Psi}(\mathbf{r}), \tag{4}$$

where

$$\tilde{\Psi}(\mathbf{r}) = \Psi(\mathbf{r}) e^{\beta E_p}. \tag{5}$$

It will be noted that $E_p$ disappears off $\tilde{\Psi}(\mathbf{r})$ while $\tilde{\Psi}(\mathbf{r}) = e^{-\beta U_e}$ for any finite $N$.

Let us set now $U_e = 0$ after the $V \to \infty$ limit and discuss the possible form of the function $\tilde{\Psi}(\mathbf{r})$ in (4). As is well known, the distribution function $F$ is an integral of motion owing to the Liouville equation. The multiplier $e^{-\beta H}$ is an integral of motion as well because the Hamiltonian $H$ is an integral of motion. Consequently the function $\tilde{\Psi}(\mathbf{r})$ should also be an integral of motion. However there are no integrals of motion dependent on the spatial coordinates alone and independent of the velocities (except for an identical constant). This is readily seen from simple considerations. The existence of such an integral would signify that upon specifying the coordinates of all particles but one we would uniquely find the coordinates of this last particle regardless of the velocities of the particles. This would signify also that changing the position of a particle would instantly affect the positions of other particles. The presence of similar rigid constraints in the system of particles under consideration is devoid of physical sense.

The absence of integrals of motion dependent on the spatial coordinates alone can be readily proven mathematically. According to the definition, an integral of motion $I(\ldots,\mathbf{r}_i,\ldots,\mathbf{v}_i,\ldots,t)$ satisfies the equation (for generality sake, allowances are made for integrals of motion that can depend upon time $t$)

$$\frac{\partial I}{\partial t} + \sum_i \mathbf{v}_i \frac{\partial I}{\partial \mathbf{r}_i} + \sum_i \frac{d\mathbf{v}_i}{dt}\frac{\partial I}{\partial \mathbf{v}_i} = 0. \tag{6}$$

In the case of $I = \widetilde{\Psi}(\mathbf{r})$, the first and last terms in (6) vanish. By virtue of the mutual independence of the coordinates and velocities the middle term will vanish for arbitrary $\mathbf{v}_i$ only if

$$\frac{\partial \widetilde{\Psi}(\mathbf{r})}{\partial \mathbf{r}_i} = 0 \qquad (7)$$

for any $i$. It immediately follows herefrom that $\widetilde{\Psi}(\mathbf{r})$ = constant.

Hence we see that upon carrying out the procedures prescribed by the method of quasi-averages we obtain, from (4), the same Gibbs distribution of (1) which yields the single particle distribution function constant in space. As a result, the method of quasi-averages cannot lead to the spatially periodic density characteristic of a crystal and does not permit one to describe the crystalline state of matter.

To investigate the crystalline state one can employ the approach in equilibrium statistical mechanics proposed in Ref. [5] (see also [4] where a presentation of the approach and of considerations leading to it is given on more detail). Ref. [5] is devoted to the quantum version of the approach and contains, in particular, derivation of a hierarchy of equations for equilibrium reduced density matrices obtained directly from the quantum mechanical Liouville equation. A characteristic feature of the approach is construction of thermodynamics compatible with the hierarchy again without use made of the Gibbs method. In the classical limit, the hierarchy obtained goes over into the well-known equilibrium Bogoliubov-Born-Green-Kirkwood-Yvon (BBGKY) hierarchy. One usually deduces the equilibrium BBGKY hierarchy leaning upon the canonical Gibbs distribution. The derivation of this last hierarchy made in [5] without recourse to the Gibbs distribution points to a wider area of its applicability, including description of the crystalline state with the inherent spatial periodicity. Ref. [4] contains also discussion of various results achieved on the basis of the approach of [5], among which are results concerning classical and quantum crystals. It may be added that the approach was employed also in studies on superfluidity of a perfect quantum crystal [6,7] and on incommensurate phases in classical crystals [8].